# Spatial and temporal circuit cutting with hypergraphic partitioning

Waldemir Cambiucci*, Regina Melo Silveira, Wilson Vicente Ruggiero
University of São Paulo, Department of Computer Engineering and Digital Systems
São Paulo, Brazil
*waldemir.cambiucci@usp.br

**Abstract**

*Quantum computing holds the promise of revolutionizing complex problem-solving by leveraging the principles of quantum mechanics. However, current noisy intermediate-scale quantum (NISQ) computers are constrained by limited qubit counts and high error rates, making it challenging to execute large quantum circuits with substantial depth, size, or width. To overcome these limitations and enhance scalability, two primary circuit cutting strategies have emerged: the gate-cut approach, which distributes circuit partitions across multiple processing units (spatial approach), and the qubit wire cut approach, which segments circuits into smaller parts for sequential execution (temporal approach). In this paper, we introduce a hypergraph-based methodology for quantum circuit cutting that is applicable to both spatial and temporal scenarios. By leveraging hypergraph theory and advanced partitioning heuristics, our approach aims to optimize circuit cutting by reducing communication overhead in spatial distributions and minimizing initialization costs in temporal sequences. We represent quantum circuits as high-level hypergraphs encompassing various combinations of quantum gates and compare partitioning heuristics such as Stoer-Wagner, Fiduccia-Mattheyses, and Kernighan-Lin for efficient graph and hypergraph partitioning. To evaluate the effectiveness of our method, we introduce a new metric called the coupling ratio, serving as a critical dimension in assessing circuit partitions. Our comparative analysis demonstrates that hypergraph partitioning enhances the efficiency of distributed quantum computing architectures by effectively balancing the trade-offs between communication and initialization costs. Specifically, heuristics like Fiduccia-Mattheyses exhibit greater flexibility and speed, making them an excellent choice for real-time circuit cutting processes in multi-QPU quantum machines. The results of our research highlight hypergraph partitioning as a pivotal step in developing a new reference architecture for quantum computers within distributed computing environments, advancing the scalability and performance of quantum computations.*

## 1. Introduction

Quantum computing holds the promise of unprecedented benefits in accelerating solutions to complex problems by harnessing the principles of quantum mechanics. However, numerous challenges hinder the application of such advantages in all scenarios. These challenges permeate the entire development stack of quantum machines [1], from the physical implementation of qubits to hardware control [2], from control software to real-world applications. This landscape delineates the computational power limitation of current machines, known as Noisy Intermediate-Scale Quantum (NISQ) devices [3], which feature few qubits, high error rates, operation latency, low gate fidelity, and short-lived states, thereby hindering the execution of large algorithms. At the current stage of quantum computers, the discipline of circuit cutting is crucial to achieve greater computational power for executing more complex algorithms.

Distributed quantum computing is defined as hardware with multiple connected quantum processing units (QPUs). These processors can be connected using direct or indirect quantum links, which create challenges for latency and provisioning. Still, communication between two QPUs can be realized using a shared pair of entangled qubits (ebits) and a teleportation protocol. In this context, entanglement sharing is at leat 10x slower than local operations inside a partition running in a single QPU. Therefore, a critical challenge is how to reduce the impact of many cuts in the original circuit before distribution among QPUs from a distributed environment.

In the literature, we found studies that explored various approaches for distributed systems, such as [4], [5], circuit distribution [6], and circuit partitioning, through the generation of smaller segments for distribution [7], [8]. These studies have laid the groundwork in the field of circuit cutting, delineating different approaches to the challenge of quantum circuit segmentation. However, it is necessary to consolidate these techniques within the context of a modular architecture of quantum computers, considering the operational costs of communication, latency, circuit coupling, and qubit dedication for communication between partitions. These are the real challenges present in the NISQ machines of various technologies for implementing current physical qubits.

In this context, two possible approaches emerge in the process of cutting quantum circuits: spatial and temporal [9].

In the spatial approach, also called the gate cut approach, we reduce the width of the original circuit and create segments with a lower number of qubits. For example, an original quantum circuit with eight qubits can be cut into two subcircuits with four qubits each in a balanced scenario. In this context, the input circuit is segmented using quantum operations or gates. When a quantum gate is cut, each participating qubit is allocated to a distinct processing unit. This demands the use of communication protocols between partitions, such as quantum teleportation, which requires the dedication of qubits in both partitions for this communication to occur. An objective function for this scenario is to reduce the communication cost involved in spatial partitioning, thereby reducing the number of generated entangled bits (ebits) or qubits dedicated to communication between partitions. In this process (spatial segmentation), the generated partitions have a smaller width; therefore, each partition has fewer qubits than the original circuit. For a balanced bipartite spatial partitioning scenario, each partition will have a width that is half the number of qubits in the original circuit.

In the temporal approach, also called the qubit-wire cut approach, we reduce the depth of the circuit so that several segments from the original circuit can be created and executed in sequence, including post-processing with classical simulations to combine the results. In this scenario, the input circuit is segmented through cuts along the qubit lines, effectively separating groups of quantum operations and enabling a deep circuit to be executed on a QPU with temporal execution constraints. In this process, the generated segments are executed serially with the challenge of initializing the qubits connected between the generated segments. In this process (temporal segmentation), the generated partitions retain the same number of qubits as in the original circuit.

Figure 1 illustrates both the spatial and temporal circuit-cutting strategies.

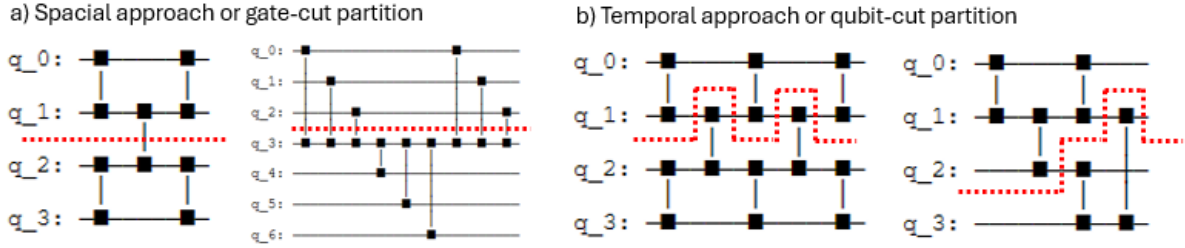

*Figure 1. Examples of Spatial (a) and Temporal (b) approach for quantum circuit cutting for static circuits. Source: created by the author (2025).*

Considering both approaches, circuit distribution among processing agents has emerged as a viable strategy for achieving greater scalability with current hardware technologies, albeit at the expense of increased control over communication among the involved agents [10]. NISQ quantum computer architectures with multiple processing units must consider additional steps of quantum circuit partitioning to facilitate the generation of subcircuits with reduced communication or initialization costs between partitions.

This paper is organized into six sections. Following this introduction, section II delves into the concept of the coupling ratio and the minimum cut of ebits. Section III introduces hypergraphic representation, which is divided into three sub-sections: gate-cut representation using a spatial approach, qubit wire-cut representation utilizing a temporal approach, and incidence matrix analysis for dual hypergraphic partitioning within the temporal approach. The methodology is detailed in section IV, outlining the approach used to address the problem. Section V presents the experiments and results, encompassing various scenarios, such as full circuits, random circuits with native gates, random circuits with independent gates, and benchmark circuits with both native and independent gates, all analysed within the context of spatial partitioning. Finally, section VI presents the conclusions based on the findings of this study.

**2. Literature Review**

Circuit cutting has become an essential technique in quantum computing, enabling the execution of large-scale quantum circuits on machines with limited qubit availability and coherence times. By dividing complex quantum circuits into smaller, more manageable segments, circuit cutting allows simulations of larger circuits beyond the native capabilities of current quantum computers. Significant advancements in this area have been achieved, focusing on enhancing efficiency [52], optimizing the partitioning process [7][8], reducing communication costs [51], and mitigating errors in noisy quantum environments [49].

Innovative methods have been introduced to optimize circuit partitioning. The Generalized Circuit Partitioning approach [39] employs a graph-based method that jointly optimizes gate and state teleportation costs, significantly reducing entanglement bit (e-bit) costs and improving time efficiency across various circuits. The Integrated Quantum Reuse and Circuit Cutting (IQRC) strategy [40] combines qubit reuse with circuit cutting to achieve a 34% reduction in the number of cuts required, thereby optimizing resource utilization and minimizing classical post-processing overhead. The Maximum Likelihood Fragment Tomography (MLFT) method [41] employs maximum likelihood estimation to minimize classical computational overhead by reconstructing the most probable distribution of the original quantum circuit output from its fragments. Error mitigation strategies, such as those explored by Majumdar and Wood [42], improve circuit cutting performance in noisy environments by addressing errors that accumulate during quantum computations.

The Rotation-Inspired Circuit Cut Optimization (RICCO) [43] method optimizes the placement of cuts within a quantum circuit to reduce the complexity of reconstructing the circuit's output, resulting in more efficient computations. The development of entanglement-based criteria for circuit fragmentation [44] helps preserve quantum correlations necessary for accurate computations by minimizing the loss of qubit entanglement during the cutting process. Additionally, the combination of virtual distillation with circuit cutting has shown promise in reducing the impact of noise on quantum computations, improving the accuracy of results from noisy quantum computers.

Graph and hypergraph methodologies have also played a crucial role in advancing circuit cutting techniques. These methods allow for the systematic partitioning of quantum circuits by representing them as graphs or hypergraphs, thus minimizing communication overhead, preserving entanglement, and optimizing resource allocation. The FitCut approach [45] transforms quantum circuits into weighted graphs and applies community detection algorithms to identify tightly connected subgraphs within the circuit. Techniques for automating the distribution of quantum circuits using hypergraph partitioning have been developed by Andrés-Martínez et al. [46], while Duncan et al. [47] employed ZX-calculus to simplify quantum circuits before fragmentation, reducing circuit complexity. FragQC [48] uses a weighted graph representation to facilitate efficient fragmentation by balancing computational load and minimizing communication between fragments. Strategies such as those proposed by Lowe et al. [49] use randomized measurements in conjunction with graph-based methods for fast and accurate circuit cutting. Davarzani et al. [37] have developed a dynamic programming algorithm that minimizes communication in distributed quantum computation by optimally partitioning quantum circuits into smaller segments.

Given these advancements, future research could explore integrating adaptive quantum circuits into existing graph and hypergraph partitioning strategies. Adaptive circuits introduce a dynamism that static graph models do not fully capture, posing a significant opportunity for further optimization in both static and dynamic quantum circuits for large-scale applications on near-term quantum hardware.

## 3. Materials and Methods

### 3.1. Full circuits as benchmarks

The full circuit is a quantum circuit characterized by the following properties.

- Complete Pairwise Interaction: For a set of n qubits $\{q_0, q_1, q_2, \ldots, q_{n-1}\}$, the circuit contains exactly one two-qubit gate between each distinct pair of qubits. That is, for every pair $(q_i, q_j)$, where:
  - $i \neq j$ and $0 \leq i, j < n$, there exists exactly one two-qubit gate $G_{i,j}$ that acts on qubits $q_i$ and $q_j$.
- Uniqueness of Gates: No two-qubit gate acts on the same pair of qubits more than once. Each gate $G_{i,j}$ is unique to its corresponding pair of qubits $(q_i, q_j)$.
- Maximal Connectivity: Each qubit $q_i$ participates in $n-1$ two-qubit gates and interacts directly with every other qubit in the circuit.

- Total Number of Gates: The total number of two-qubit gates in the circuit is given by C(n,2) = n(n−1)/2.
- Optional Single-Qubit Gates: While defining feature of a full circuit is the complete set of two-qubit interactions, single-qubit gates may also be applied to qubits as needed for specific algorithms.

The set of properties for the full circuits is as follows:

- Complete entanglement: The full circuit ensures the maximum entanglement potential, as each qubit is directly entangled with every other qubit through a two-qubit gate.
- Symmetry: The circuit is symmetric with respect to qubit interactions; all qubits play an equivalent role in terms of connectivity.
- Scalability: As $n$ (number of qubits in the circuit) increases, the number of required two-qubit gates grows quadratically, which may impact practical implementation on hardware with connectivity constraints.

Using full circuits, it is possible to explore the maximum impact of minimum-cut heuristics for different combinations of qubits in quantum circuits during the partitioning process.

**3.2. Coupling ratio and min.cut of ebits**

For any quantum circuit, we can calculate different dimensions as depth (d), that is, the number of layers or time steps required to execute all gates in the circuit in sequence, width (w) is the maximum number of qubits actively involved in any single layer or time step of the circuit, and size (s) is the total number of gates or operations within the circuit. In this study, we explored a new dimension called the coupling ratio (Cr).

We can calculate the coupling ratio dimension in terms of the number of quantum gates present in the circuit against its width or the number of qubits. To achieve this, we can use another dimension called the coupling base (Cb).

Cb represents the number of unique binary quantum gates possible for a full circuit, considering the same number of qubits (n). Therefore, for a full circuit with n qubits, we can calculate Cb as a combination problem (C function) using Eq. (1):

$$Cb = C(n, binary\ gates) = C(n, 2) = \frac{n \times (n-1)}{2} \quad (1)$$

For a full circuit with four qubits (n=4), we calculated Cb = 6. For n=8, Cb = 28, etc.. Figure 2 illustrates the layout of the full circuits and their Cb dimensions.

Based on this value, we can define a new dimension called the Coupling Ratio (Cr), represented by Eq. (2):

$$Cr = \frac{total\ number\ of\ n-qubits\ gates}{Cb} \quad (2)$$

Therefore, for an ordinary circuit with n qubits and x number of n-qubit gates, Cr can be calculated using the quivalent Cb for the same number of qubits and the total number of binary gates presented in the circuit. For Cr < 1, the target circuit presents a lower interconnectivity among the qubits, while Cr > 1 represents a quantum circuit with higher interconnectivity.

This value can help guide the circuit-partitioning process, thereby influencing the impact of circuit cutting on the target circuit. Figure 2 illustrates different representations of a full circuit.

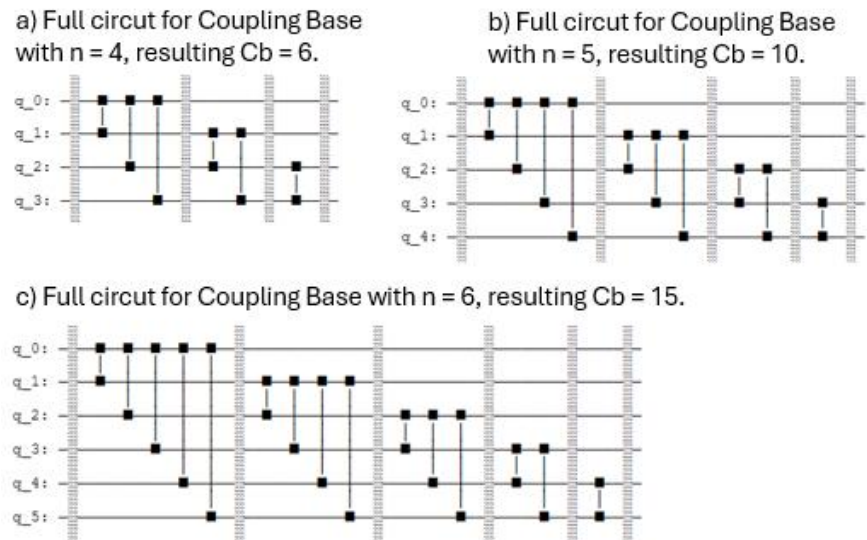

*Figure 2. Examples of full circuits to calculate the Coupling Base (Cb) with n=4, 5, and 6 (number of qubits). Source: Created by the author (2025)*

Although full circuits present all sets of unique quantum n-qubit gates, it is possible to calculate max. base number of gate cuts for a scenario of central bipartition, when we cut the circuit right in the middle, creating two partitions with a similar number of qubits in the spatial approach.

We can consider figure 3, with the graphic for the Coupling Base (Cb) and the number of ebits generated by the bipartition process for full circuits from 4 to 120 qubits. Although we do not have duplicated gates, the full circuit is at the maximum depth, and the binary cut is in the minimum cut of ebits between partitions.

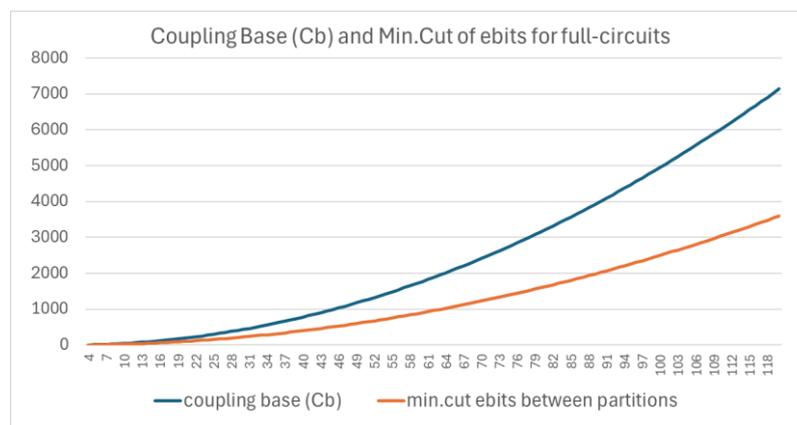

*Figure 3. Graphic for Coupling Base and Min.Cut ebits for full circuits from 4 to 120 qubits. Source: Created by the author (2025)*

From figure 3, we can see a significant connection between the coupling base graphic and the number of ebits cuts for full circuits from four to 120 qubits. Considering them as a standard metric for the number of ebits, the impact of circuit cutting for other algorithms can be compared against these graphics, creating a metric for a future resource estimator around circuit cutting.

## 3.3. Hypergraphic representation

A graph is a mathematical model represented by G = (V, E), where V denotes the vertices and E denotes the edges, representing binary relationships. In a formal definition, a graph G = (V,E), where V is a set and E is a subset of subsets of size two of V. A hypergraph is a mathematical model represented by H = (V, E), where V refers to vertices (also called nodes) and E refers to hyperedges, representing non-binary relationships, allowing a hyperedge to wrap around more than one element [11] or multiple connections among multiple vertices. In a form definition, a hypergraph H = (V,E), where V is a set and E is a family of subsets of V.

A graph G in the context of quantum circuits is a structure where vertices represent qubits and edges represent binary quantum gates, such as CNOT, CZ, CP, and SWAP. In turn, a hypergraph H can be used as a generalization of graphs and is applicable to more complex scenarios in quantum circuits, covering multi-qubit gates such as CCNOT or CSWAP, tripartite entanglement correlations, quantum error correction codes, or complex state representations. Therefore, a primal hypergraph H is a structure in which vertices represent qubits and hyperedges represent quantum gates for two or n qubits. We can visualize hypergraphs where vertices are points and edges are lines enclosing points in the layout.

Graph and hypergraph partitioning are optimization problems that focus on dividing a graph or hypergraph into multiple partitions while minimizing specific objective functions. For graph partitioning, a variety of heuristics can be employed, including Spectral Bisection [12], kernighan-Lin (KL) algorithm [13], Metis [14], Diffusion-Based Methods [15], Greedy Algorithms [16], Genetic Algorithms [17], Stoer-Wagner (SW) algorithm [18], and Force-Directed Methods [19]. In the context of hypergraph partitioning, techniques such as Recursive Bisection [20], K-way Partitioning [21], the Fiduccia-Mattheyses (FM) algorithm [22], Multilevel Hypergraph Partitioning [23], Hypergraph Balancing [24], and Min-Cut Algorithms [25] are utilized. Hybrid Methods [26].

These methods provide various strategies for partitioning, each with its own advantages in terms of efficiency and effectiveness in minimizing the objective functions associated with the partitioning problem.

In figure 4, we present an example of a quantum circuit and its representation in a primal hypergraph before the partitioning process.

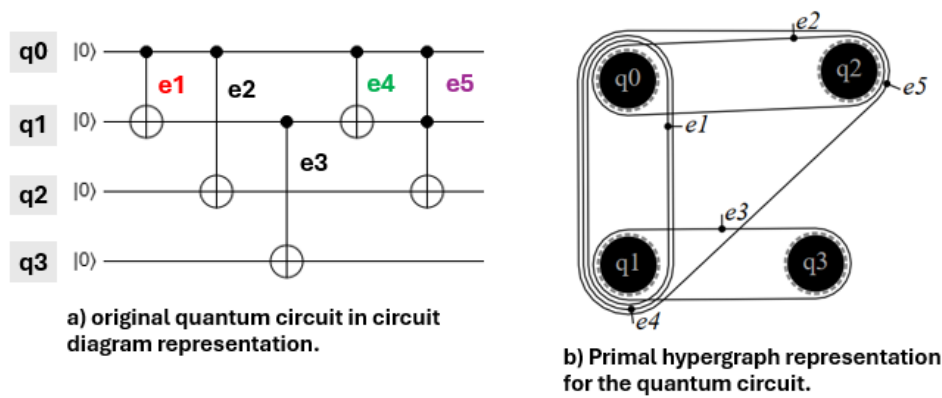

*Figure 4. Quantum circuit (a) and primary hypergraphic representation (b). Source: Created by the author (2025)*

The implementation of hypergraph partitioning enhances the segmentation process and offers significant benefits for circuit cutting. These advantages include the following.

Improved Results for Large and Complex Circuits: Hypergraph partitioning is well suited to handle wide and intricate circuits, leading to more effective segmentation compared to traditional methods.

Better Representation of Multi-Qubit Operations: It provides superior adherence in representing operations that involve multiple qubits and complex interconnections, thereby capturing the nuances of quantum gate interactions more accurately.

Efficient Optimization of Communication: By minimizing the communication between distant qubits, hypergraph partitioning reduces the partition execution time and improves the overall performance of the circuit.

It is important to note that different techniques can be applied to create circuit representation in a hypergraph layout. This flexibility allows us to guide hypergraph partitioning heuristics by focusing on either qubit wire cuts or gate-cut approaches. By selecting an appropriate method, we can optimize the partitioning process to suit the specific requirements of the quantum circuit in question.

### 3.3.1. Hypergraphic representation for gate-cut (spatial approach)

From the input circuit in its circuit diagram representation, we can use Algorithm 1 to create a hypergraph structure of the circuit, where vertices represent qubits and hyperedges represent n-qubit quantum gates.

*ALGORITHM 1. To translate a quantum static circuit diagram into a hypergraphic representation, we use H=(V,E).*

```
def create_hypergraph_from_quantum_circuit(circuit_diagram):
    # Initialize an empty hypergraph.
    hypergraph = {'vertices': [], 'hyperedges': []}

    # Iterate through each gate in the quantum circuit diagram.
    for gate in the circuitdiagram:
        qubits = gate.qubits
        Operation = gate operation

        # Qubits are added as vertices to the hypergraph.
        for qubit in qubits:
            if qubit is not in the hypergraph['vertices']
                Hypergraph ['vertices']. append(qubit)

        # Add a hyperedge representing the quantum gate.
        hypergraph['hyperedges']. append({'operation': operation, 'qubits:' qubits})
    return hypergraph
```

The returned hypergraph can be manipulated as an incidence matrix, where we identify the vertices belonging to each hyperedge E of the hypergraph H. This set is important in the partitioning heuristic, whereas we move the vertices between partitions. In a Fiduccia-Mattheyses-based heuristic, the objective is to generate two balanced partitions with an equal number of vertices in each partition, thereby reducing the number of hyperedges cut between the partitions.

The impact of this objective in the context of quantum circuits is the reduction in communication links required between partitions, thus minimizing the use of quantum teleportation between partitions during execution across multiple QPUs, as shown in figure 5.

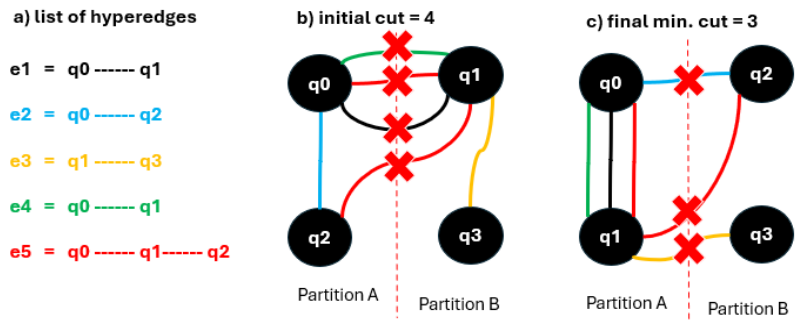

*Figure 5. Hypergraphic partitioning by Fiduccia-Mattheyses was applied for the hypergraph representation of a quantum circuit. In (a), we have the list of hyperedges; in (b), we start the heuristic with an initial random partition; in (c), we finish with a better min.cut. Source: Created by the author (2025)*

This primal hypergraph works well for spatial circuit cutting scenarios, where the hypergraph representation is directly used as the input for a partitioning heuristic.

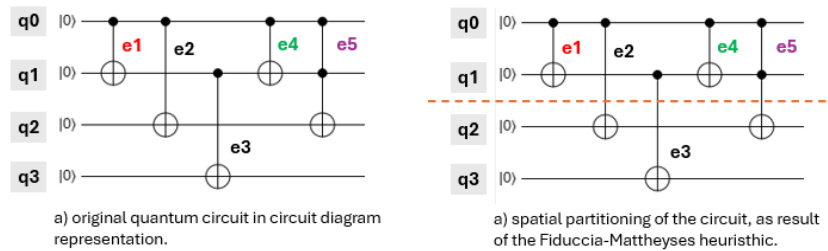

*Figure 6. Hypergraphic partitioning results apply to quantum circuits. Source: Created by the author (2025)*

Using this spatial approach and hypergraph representation with our benchmark circuit, we end the process with three cuts, grouping qubits q0 and q1 in one partition and qubits q2 and q3 in another partition, as shown in figure 6.

For temporal circuit cutting scenarios, where we maintain the circuit width and cut along the qubit lines in the circuit diagram, we can apply a new approach based on the dual representation of the hypergraph.

### 3.3.2. Hypergraphic representation for qubit wire-cut (temporal approach)

In the temporal approach to circuit segmentation, we can use equivalent dual hypergraph generated from the input circuit. In this approach, the original vertices (qubits) become hyperedges in the dual hypergraph, whereas the hyperedges from the original circuit become vertices.

The dual hypergraph (H*) of a given primal hypergraph (H) is another hypergraph that captures complementary information regarding the original hypergraph [27]. In the dual hypergraph, the roles of vertices and hyperedges are reversed: vertices in the original hypergraph become hyperedges in the dual hypergraph and hyperedges in the original hypergraph become vertices in the dual hypergraph. The relationship between a hypergraph and its dual hypergraph can be understood as a kind of "duality" or "reciprocity." It provides an alternative perspective on the same underlying structure, highlighting different aspects of connectivity and relationships among elements [28]. In the context of quantum circuit cutting, working with dual hypergraphs can explore the minimum cut of wired qubits in the circuit, reducing the impact of circuit partitioning on initialization needs among many serialized segments of the circuit.

To accomplish this, we have an additional step: translating hypergraph H into a dual hypergraph H*. Figure 7 shows an example of primal and dual hypergraphs for a quantum circuit with four qubits, using the same circuit from figure 4.

Therefore, for the hypergraph H=(V,E) of order n and size m, consider the hypergraph H*=(E,X), where the vertices are the edges of H, and whose edge family X is formed as follows:

- X is the family of sets x1, x2, x3,...,xn, where xi is the set of all edges in E inciding on vertex vi in H.

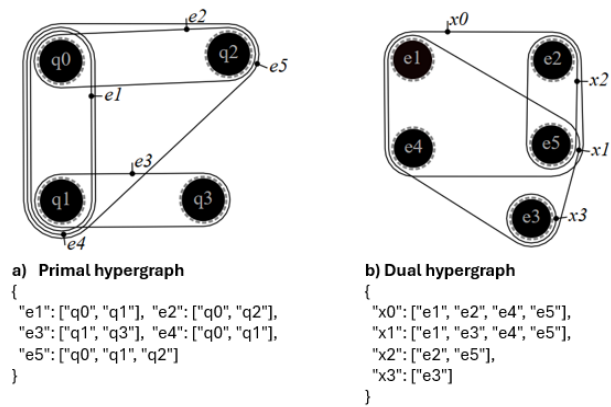

Figure 7. Transformation of primal hypergraph H (a) into its equivalent dual hypergraph H*(b). Source: Created by the author (2025)

Therefore, informally, we transform vertex vi into edge xi and transform the edges into vertices. As a direct result, when we use the dual hypergraphic representation of a quantum circuit for the temporal partitioning process, we reduce the impact of the initialization of qubits in subsegments.

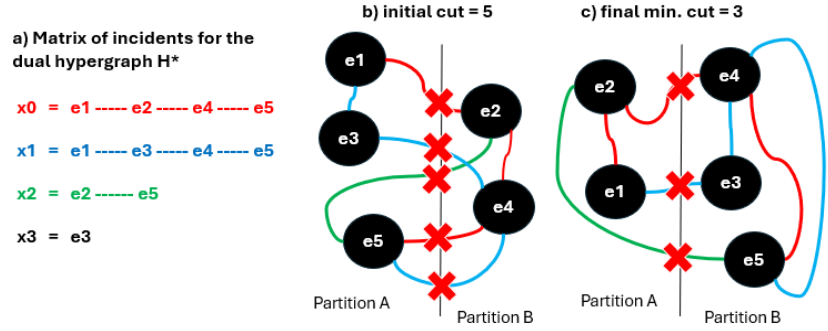

*Figure 8. Dual hypergraphic partitioning using Fiduccia–Mattheyses In (a), we have a list of hyperedges; (b) starting with a random partition; and (c) finishing with three qubit-wire cuts in the temporal approach. Source: Created by the author (2025)*

In figure 8, we illustrate the initial and final stages of hypergraphic partitioning using Fiduccia-Mattheyses, using the dual hypergraphic representation for the benchmark circuit with four qubits. It is possible to see a reduction in the number of links cut at the end of this process, which has a significant impact on the initialization of segments in the temporal approach.

Algorithm 2 presents the pseudocode for transforming hypergraph H into the equivalent dual hypergraph H*.

*ALGORIHM 2 – Algorithm for returning the dual hypergraph H* from the primal hypergraph H representing the quantum circuit.*

```
def create_dual_hypergraph_from_hypergraph(hypergraph)
    # Initialize an empty dual hypergraph.
    dual_hypergraph = {'vertices': [], 'hyperedges': []}

    # Iterates through each vertex in the original hypergraph
    for vertex in the hypergraph['vertices']:
        # The vertex is added as a hyperedge to the dual hypergraph.
        dual-hypergraph ['hyperedges']. append({'qubit': vertex})

    # Iterates through each hyperedge in the original hypergraph
    for hyperedge in the hypergraph['hyperedges']
      ' qubits' = hyperedge['qubits']
      operation = hyperedge[operation']

      # Add qubits as vertices to the dual hypergraph.
      for qubit in qubits:
         if qubit is not in dual_hypergraph['vertices']
            dual-hypergraph ['vertices']. append(qubit)

      # Add a hyperedge representing the quantum gate.
      ypergraph['vertices']. append({'operation': operation, 'qubits:' qubits})
    return dual_hypergraph
```

Finally, this configuration represents a temporal cutting approach in which the original circuit can be serialized in a monolithic QPU system, as illustrated in figure 9. It is possible to observe the two segments created at the end of this partitioning process using a temporary approach.

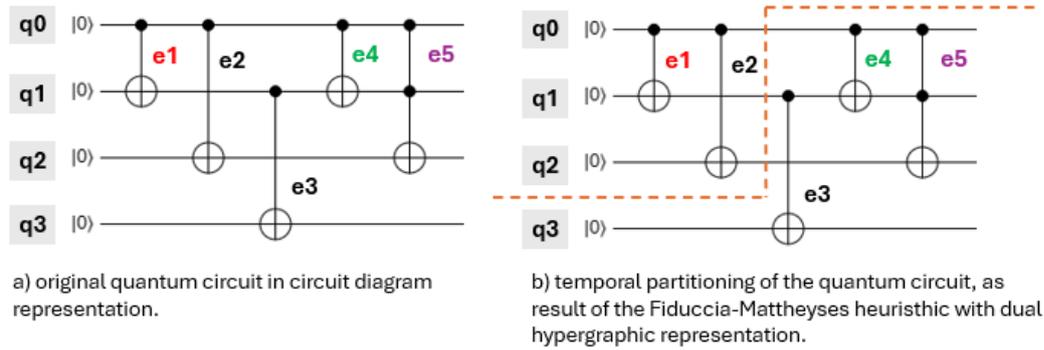

a) original quantum circuit in circuit diagram representation.

b) temporal partitioning of the quantum circuit, as result of the Fiduccia-Mattheyses heuristhic with dual hypergraphic representation.

*Figure 9. The temporal approach was applied to the quantum circuit, creating two segments for execution. Source: Created by the author (2025)*

### 3.3.3. Incidence matrix for dual hypergraphic partitioning in temporal approach

Finally, for every primal hypergraph created for experiments using the spatial approach, we can create an equivalent dual hypergraph as input in the temporal cutting approach. To achieve this, a primal hypergraph can be represented by an incidence matrix to support a new partitioning process.

Consider the quantum circuit in figure 10 with six qubits (width=6) and 10 gates (size=10).

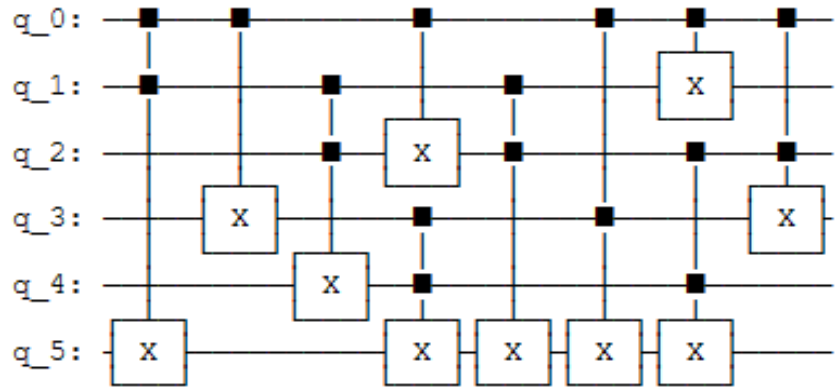

*Figure 10. Quantum circuit example with binary ('cx') and ternary ('ccx') gates Source: Created by the author (2025).*

For this quantum circuit, we can create a primal hypergraph H with different gates in the circuit, as listed in Table 1, which presents the incidence matrix representing the primal hypergraph.

An incidence matrix is a mathematical representation used in graph and hypergraph theories to describe the relationships between vertices (nodes) and edges (hyperedges) in a graph or hypergraph.

In the case of hypergraph, the incidence matrix describes the nodes that are incident to each hyperedge. Each row of the incidence matrix corresponds to a hyperedge, and each column corresponds to a node. A '1' in a cell indicates that the corresponding node is incident to the hyperedge, whereas a '0' indicates that it is not.

Similarly, for the dual hypergraph, each row corresponds to a node and each column corresponds to a hyperedge.

TABLE 1. *Incidence matrix for the primal hypergraph, for example with 6 nodes and 10 hyperedges. For this incidence matrix, num.lines = size and num.colunms = width for the circuit. Source: created by the author (2025).*

|  |  | nodes | | | | | |
|---|---|---|---|---|---|---|---|
|  |  | 1 | 2 | 3 | 4 | 5 | 6 |
| hyperedges | 1 | 1 | 1 | 0 | 0 | 0 | 1 |
|  | 2 | 1 | 0 | 0 | 1 | 0 | 0 |
|  | 3 | 0 | 1 | 1 | 0 | 1 | 0 |
|  | 4 | 1 | 0 | 1 | 0 | 0 | 0 |
|  | 5 | 0 | 0 | 0 | 1 | 1 | 1 |
|  | 6 | 0 | 1 | 1 | 0 | 0 | 1 |
|  | 7 | 1 | 0 | 0 | 1 | 0 | 1 |
|  | 8 | 1 | 1 | 0 | 0 | 0 | 0 |
|  | 9 | 0 | 0 | 1 | 0 | 1 | 1 |
|  | 10 | 1 | 0 | 1 | 1 | 0 | 0 |

Similarly, the equivalent dual hypergraph can be represented by a new incidence matrix, with rows representing nodes and columns representing hyperedges. This simple way to represent circuits for the partitioning process brings a huge benefit to quantum circuits with a higher number of quantum gates, especially in the scenario of temporal cutting for a monolithic QPU scenario, with restrictions for circuits with higher values of depth. Table 2 illustrates the incidence matrix for dual hypergraphic as discussed.

TABLE 2. *Incidence matrix for a dual hypergraph, equivalent to an example with six nodes and ten hyperedges. For this incidence matrix, num.lines = circuit width and num.colunms = circuti size. Source: Created by the author (2025)*

|  |  | nodes | | | | | | | | | |
|---|---|---|---|---|---|---|---|---|---|---|---|
|  |  | 1 | 2 | 3 | 4 | 5 | 6 | 7 | 8 | 9 | 10 |
| hyperedges | 1 | 1 | 1 | 0 | 1 | 0 | 0 | 1 | 1 | 0 | 1 |
|  | 2 | 1 | 0 | 1 | 0 | 0 | 1 | 0 | 1 | 0 | 0 |
|  | 3 | 0 | 0 | 1 | 1 | 0 | 1 | 0 | 0 | 1 | 1 |
|  | 4 | 0 | 1 | 0 | 0 | 1 | 0 | 1 | 0 | 1 | 1 |
|  | 5 | 0 | 0 | 1 | 1 | 1 | 0 | 0 | 0 | 1 | 0 |
|  | 6 | 1 | 0 | 0 | 0 | 1 | 1 | 1 | 0 | 0 | 0 |

### 3.4. Methodology

To explore the impact of hypergraphic representation in primal and dual hypergraph formats, we used a combination of experiments with full circuits, random circuits, and benchmark circuits, ranging from 4 to 120 qubits.

A full circuit is considered a standard circuit with a full combination of binary gates, CZ or CX, among all qubits, with no duplicated gates. In this way, we can compare different metrics and the impact of circuit partitioning on the minimum and maximum number of gates affected by cuts. A random circuit is created automatically by a tool that is twice as deep as wide for a specific number of qubits and has random combinations of quantum gates from a specific gate set. A benchmark circuit is selected from a list of known algorithms, ranging in the number of qubits, such as Grover's algorithm (GROVER) [33], Quantum Phase Estimation (QPE) [34], Quantum Fourier Transformation (QFT) [35], Variational Quantum Eigensolver (VQE) [36], and Quantum Approximation Optimization Algorithm (QAOA) [37].

We used the Munich Quantum Toolkit Benchmark Library [30] as a tool for the creation of random and benchmarking circuits for different variables and setups. The complete setup of the circuits used in this methodology is as follows.

- Full circuits were created with a full set of binary gates for all qubits with no duplicated operations.
- Random circuit using a set of native gates.
- Random circuit using a set of independent gates.
- Benchmark circuit with a set of native gates.
- Benchmark circuit with set of independent gates

As benchmark circuits for the experiments, we selected the following circuits as representative algorithms for different applications.

- Grover's algorithm (No Ancilla)[33]: a well-known quantum algorithm that helps find a specific target quantum state using an oracle. In this version, the MQT Bench implements the oracle using a multi-controlled Toffoli gate over all input qubits. Notably, this version did not require ancilla qubits during its execution.
- The Quantum Approximation Optimization Algorithm (QAOA) [37] is a well-known algorithm belonging to the variational quantum algorithm class. It is designed to tackle optimization problems and can be adjusted using these parameters. In one example, it was used to solve a max-cut problem.
- The Quantum Fourier Transformation (QFT) [35] is the quantum equivalent of the discrete Fourier transformation. They play a crucial role as fundamental building blocks in various quantum algorithms.
- Quantum Phase Estimation (QPE) – Exact [34] was utilized to estimate the phase of a quantum operation. This is a critical component in many quantum algorithms. In an exact scenario, the phase applied can be precisely represented by the number of qubits involved.
- The Variational Quantum Eigensolver (VQE) [36] is another renowned algorithm within the variational quantum algorithm class. Like QAOA, it is parameterizable and offers various choices for an ansatz function.

With a set of native gates, we are creating quantum circuits for a specific platform, such as an ibm machine, using Qiskit. In this scenario, the generated circuit is using only a small set of basic gates available in the platform, such as 'rz, ' 'sx, ' 'x, ' 'cx,' and terminals as 'measure' and 'barrier'.

In this case, we don´t have 3-qubit gates, but only unitary gates and binary gates, so the partitioning process can be performed through graphical heuristics such as Stoer Wagner (SW) [18] or Kernighan-Lin (KL) [13].

The generated hypergraph has hyperedges with only two vertices. For native gates, it is possible to create a different set of circuits by considering optimization to reduce the number of operations. With the Qiskit platform, it is possible to create circuits with optimization levels of 0, 1, 2, and 3, creating different numbers of gates for the same circuit and ranging the depth for each circuit.

When we work with a set of independent gates, we create quantum circuits for a universal set of gates. In this scenario, the generated circuit can use a bigger set of operations, including:

- unitary gates, such as 'h', 'p', 'x', 'y', 'z', 'rx', 'rz', 's', 't'
- binary gates such as 'cx', 'cp', 'swap', 'ch', 'cz'
- and 3-qubits gates, as 'ccx' (controlled-controlled-x or Toffoli gate [31]) and 'cswap' (controlled-swap or Fredkin gate [32]).

In this case, the partitioning process cannot be performed using graphical heuristics such as Stoer Wagner[18] or Kernighan-Lin [13]. Instead, heuristics such as Fiduccia-Mattheyses (FM) [22] are the best choice for this scenario, working with a generated hypergraph with hyperedges containing three vertices and many times during the circuit. For this set of circuits, we do not have the option to optimize the circuit, so we use a complete set of gates for the unitary, binary, and 3-qubits gates set.

For instance, on platforms like IBM's Qiskit, the transpilation process breaks down complex gates like 'ccx' (Toffoli gate) into native gates from the basic set. Consequently, during transportation and gate decomposition, the resulting circuit becomes larger than the original one.

Indeed, it is still possible to apply the KL heuristic for circuit partitioning with independent gates when working with a bipartite graph format, when expanded vertices are added to represent hyperedges. In this approach, we have more vertices in the final graph representation; however, it is possible to work with graph partitioning supported by KL or SW when edges have only two possible vertices.

## 4. Experiments and results

We conducted multiple experiments using hypergraphic partitioning and KL and FM heuristics on our three sets of circuits: full circuits, random circuits, and benchmark circuits. These experiments aimed to assess the effect of the bipartite approach on the communication cost between Quantum Processing Units (QPUs) in a distributed system.

To compare the results, we measured the number of cuts made by a bipartite process at the midpoint of the circuit without reorganizing qubits between partitions and compared the results with those from hypergraphic heuristics. Therefore, it was possible to measure the creation of a standard number of entanglement bits (ebits) between partitions or QPUs and the cost reduction using hypergraphic methods when applied to spatial partitioning.

### 4.1. Experiments with full circuits for spatial partitioning

Table 3 presents the results from the experiments with Full Circuits from 4 to 16 qubits.

For each specific width (num.qubits), we have equal values for depth (d), size (s), and coupling base (Cb); for a full circuit, we have all possible combinations of n-qubit gates between qubits, with no duplicated gates in the circuit.

TABLE 3. *Experiments with full circuits from 4 to 12 qubits for balanced bipartitions (k=2) using the KL heuristic resulted in a min.cut of gates. The depth, size, and Cb were the same for the full circuit.*

| width | depth | size | Cb | mid. cut | min. cut KL | bipartite segments (k=2) |
|---|---|---|---|---|---|---|
| 4 | 6 | 6 | 6 | 4 | 4 | ({1, 2}, {0, 3}) |
| 5 | 10 | 10 | 10 | 6 | 6 | ({3, 4}, {0, 1, 2}) |
| 6 | 15 | 15 | 15 | 9 | 9 | ({0, 4, 5}, {1, 2, 3}) |
| 7 | 21 | 21 | 21 | 12 | 12 | ({0, 2, 6}, {1, 3, 4, 5}) |
| 8 | 28 | 28 | 28 | 16 | 16 | ({0, 3, 6, 7}, {1, 2, 4, 5}) |
| 9 | 36 | 36 | 36 | 20 | 20 | ({8, 2, 3, 4}, {0, 1, 5, 6, 7}) |
| 10 | 45 | 45 | 45 | 25 | 25 | ({3, 4, 5, 6, 7}, {0, 1, 2, 8, 9}) |
| 11 | 55 | 55 | 55 | 30 | 30 | ({2, 4, 6, 9, 10},{0, 1, 3, 5, 7, 8}) |
| 12 | 66 | 66 | 66 | 36 | 36 | ({3, 5, 8, 9, 10, 11},{0, 1, 2, 4, 6, 7}) |

Evidently, for a complete circuit, the number of cuts in the bipartition method is always at its maximum. This is because all qubits are engaged in the binary gates throughout the circuit. The same result is generated with the KL and FM heuristics when using the primal hypergraphic representation for each full circuit from the experiment. Considering this behavior from full circuits, figure 11 represents the primal and dual hypergraphic for a full circuit with four qubits, so we can check the completeness of the connections among qubits.

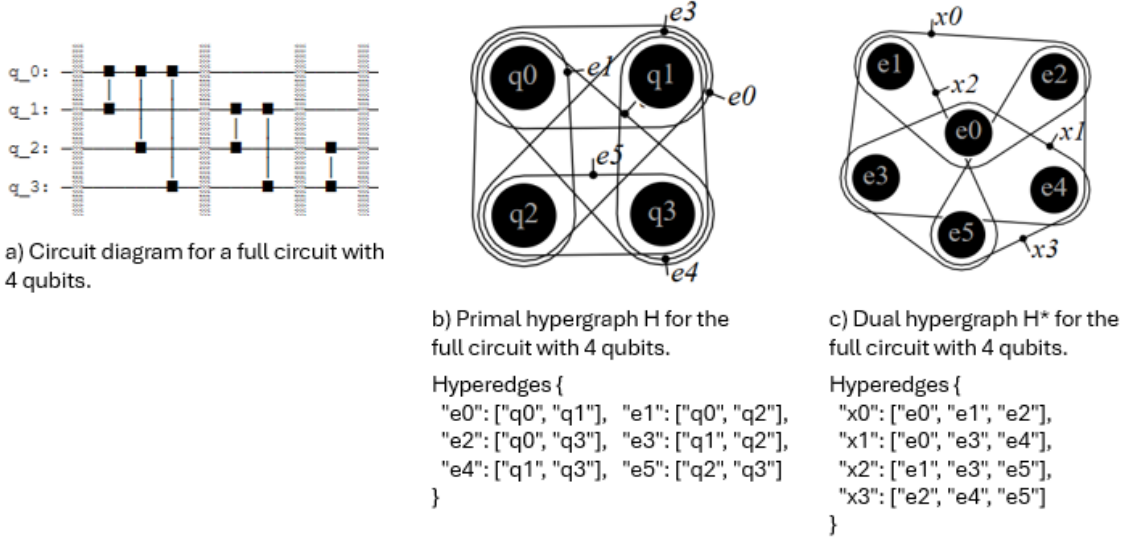

*Figure 11. Primal (b) and dual (c) hypergraphic for a full circuit (a) with four qubits. Source: Created by the author (2025)*

## 4.2. Experiments with random circuits and native gates for spatial partitioning

In table 4, we present the results of experiments with random circuits and native gates on the Qiskit platform. For this scenario, we still used hypergraphic with unitary and binary gates only.

TABLE 4. *Experiments with random circuits for native gates, from 4 to 12 qubits, for balanced bipartitions (k=2) using the KL heuristic and optimization level 0, resulting in a min-cut of gates.*

| width | depth | size | Cb | mid. cut | min. cut KL | bipartite segments |
|---|---|---|---|---|---|---|
| 4 | 91 | 168 | 6 | 17 | 14 | ({0, 3}, {1, 2}) |
| 5 | 98 | 228 | 10 | 26 | 16 | ({2, 4}, {0, 1, 3}) |
| 6 | 95 | 261 | 15 | 32 | 15 | ({0, 3, 5}, {1, 2, 4}) |
| 7 | 125 | 398 | 21 | 43 | 27 | ({2, 4, 6}, {0, 1, 3, 5}) |
| 8 | 142 | 493 | 28 | 46 | 34 | ({2, 3, 4, 7}, {0, 1, 5, 6}) |
| 9 | 147 | 607 | 36 | 62 | 42 | ({8, 1, 4, 5}, {0, 2, 3, 6, 7}) |
| 10 | 156 | 656 | 45 | 77 | 55 | ({2, 3, 5, 8, 9}, {0, 1, 4, 6, 7}) |
| 11 | 247 | 957 | 55 | 128 | 97 | ({2, 3, 4, 5, 7}, {0, 1, 6, 8, 9, 10}) |
| 12 | 329 | 1219 | 66 | 163 | 101 | ({0, 1, 2, 4, 6, 7}, {3, 5, 8, 9, 10, 11}) |

Following table 4, we repeated the experiment ranging from 4 to 120 qubits with each random circuit for native gates and still applied four sets of optimizations with levels 0, 1, 2, and 3. Figure 12 presents the performance of hypergraphic partitioning for different levels of optimization used with each random circuit.

The graph presents the % of reduction in the number of ebits created by a spatial bipartition, using a hypergraphic approach, against the number of ebits generated by a circuit cutting central in the original circuit. The experiment presented the same behavior for different levels of optimization in the circuit, and the gain from hypergraph partitioning gradually decreased from 40% compared to the central cut to 10% for benchmark circuits of larger widths (up to 120 qubits).

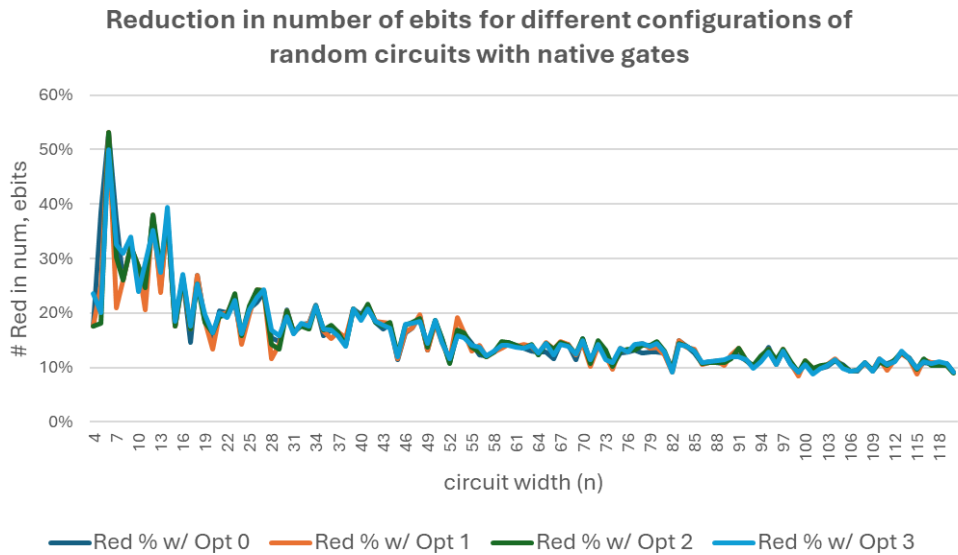

*Figure 12. Impact of hypergraphic partitioning using KL heuristic for random circuits with native gates and different levels of optimization in the qiskit platform. Source: Created by the author (2025).*

Here, we check the first important result for the impact of quantum circuit partitioning: even for a random number of gates, the reduction in the cost of communication produced by graph partitioning reduces a higher number of qubits in a larger width.

## 4.3. Experiments with random circuits and independent gates for spatial partitioning

As shown in figure 13, we conducted experiments with random circuits and independent gates. Here, we consider the hypergraphic partitioning heuristic with Fiduccia–Mattheyses for a full hypergraphic representation. We also run experiments with the Kernighan-Lin method, but with a bipartite graph representation of the circuit, so we have additional vertices called 'connectors' to support hyperedges as 'ccx' and 'cswap.' With this approach, we can still observe the same effect of partitioning while increasing the number of qubits.

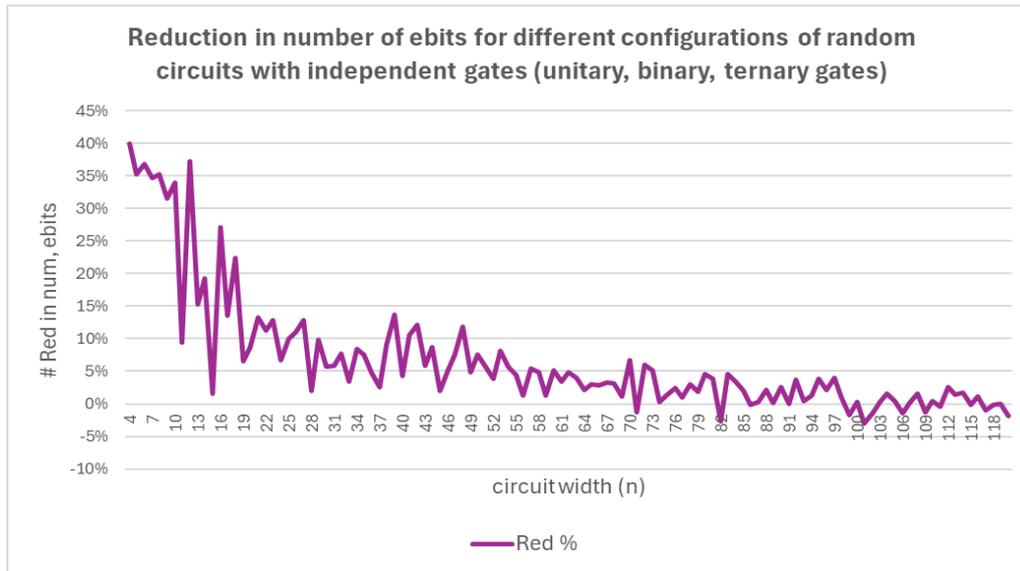

*Figure 13. Impact of hypergraphic partitioning using the KL heuristic for random circuits with independent gates for circuits ranging from 4 to 120 qubits. Source: Created by the author (2025)*

Following the same approach as the first experiment, we can see the impact of larger circuits for circuit partitioning in both scenarios of graph and hypergraphic heuristics. We see a reduction in % for the number of ebits created by the hypergraphic bipartition, compared to the central cutting, with no reorganization of qubits between partitions.

In this scenario, two observations are made: hypergraphic partitioning offers better performance for the reduction of communication between partitions, starting with 40% less ebits, and decreasing this performance for circuits with a larger number of qubits for width, until 120 qubits. For random circuits using native gates or independent gates, the circuit cutting reaches a threshold of 10% for a higher number of circuit widths, for both heuristics such as graph and hypergraph partitioning. For every experiment, the graphic for the reduction of generated ebits created by hypergraphic partitioning was better than the naive method of central cutting.

These results will be important for future exploration of circuit cutting strategies, which demand additional strategies to optimize the communication between partitions for more complex circuits.

## 4.4. Experiments with benchmark circuits and native gates for spatial partitioning

In figure 14, we present experiments with benchmark circuits and native gates on the Qiskit platform. For this scenario, we are using hypergraphic with unitary and binary gates only, while in native gates the operation sets have only one binary gate, the 'cx.' We tested circuits based on the templates of QAOA, QFT, VQE, AE, GROVER, and QPE, ranging the number of qubits between 4 and 120, and optimization levels 0,1,2, and 3 for the Qiskit platform, reducing the depth for each scenario.

For this set of experiments, circuits such as VQE and QAOQ did not demonstrate benefits from hypergraphic partitioning against the mid-cut approach. However, it is possible to observe the same behavior for the reduction in ebits created by the partitioning process. The shape of this graphic has implications for the impact of complex circuits in the circuit-cutting scenario.

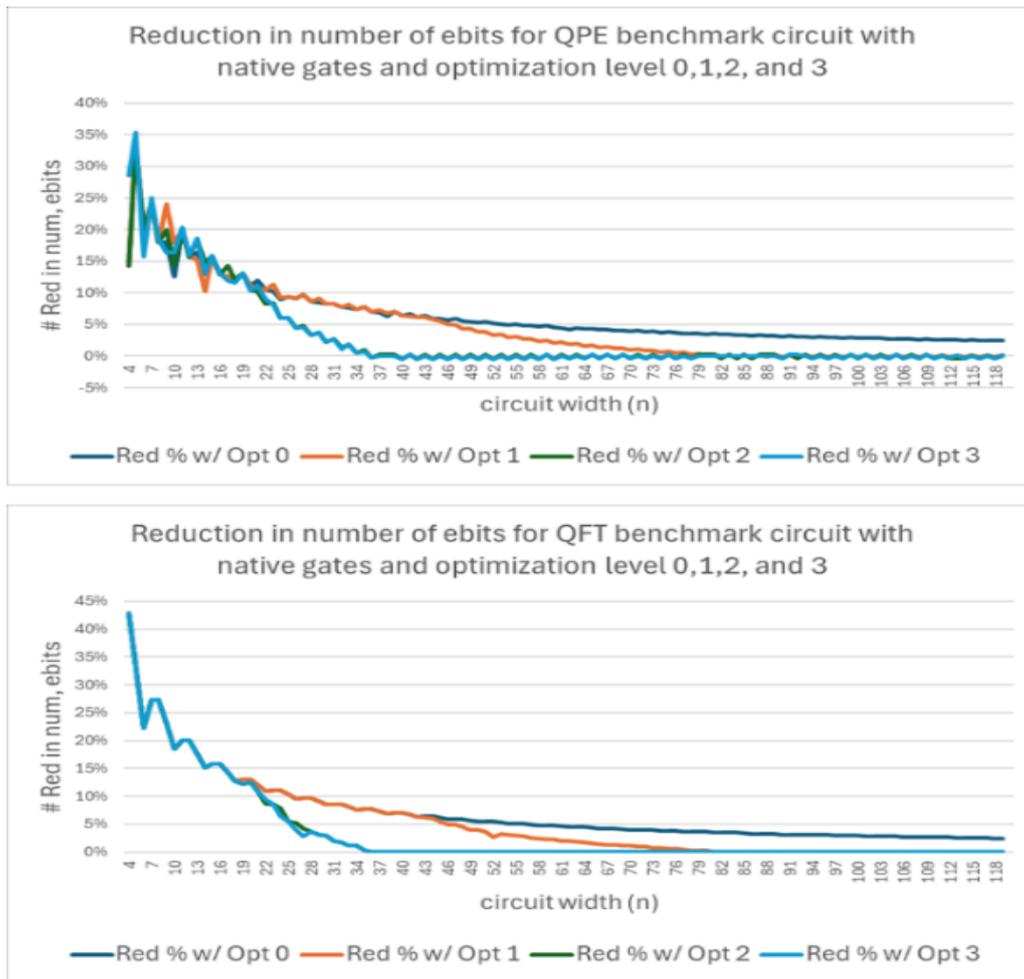

Figure 14. Impact of hypergraphic partitioning using KL heuristic for benchmark circuits with native gates Source: Created by the author (2025).

In the graph presented in figure 14, we can observe a similar behavior for the reduction in performance for the hypergraphic partitioning approach for benchmark circuits with larger numbers of qubits in width. However, for the QFT and QPE benchmark circuits, the benefits of the hypergraphic approach presented a stronger decrease, reducing to no benefits compared to the central cutting approach.

This is a relevant observation, indicating that for circuits with intensive use of binary and ternary gates, following standards such as QFT and QPE, the hypergraphic partitioning process can offer similar results for larger circuits, compared to central cutting or naïve methods, when we cut the circuit in the middle, with no reorganization of qubits during this process.

## 4.5. Experiments with benchmark circuits and independent gates for spatial partitioning

Following the series of experiments, we conducted sessions using benchmark circuits and independent gates. Here, we consider hypergraphic partitioning heuristics such as Fiduccia-Mattheyses and Kernighan-Lin applied for bipartite graph representation to support 3-qubits gates operations over the circuit and hyperedges from the primal hypergraphic representation. Working with independent gates, we did not apply any optimization step as we did with the native gate before. It was possible to observe the same shape for the impact of hypergraph partitioning for different versions of benchmark circuits when the benefit of the reduction in the number of ebits created during the partitioning process decreased for larger circuits.

In figure 15, we present the % of reduction of a hypergraphic partitioning process, obtained with the number of ebits created by the hypergraphic method, against the number of ebits created by the central cut approach, for the same circuit.

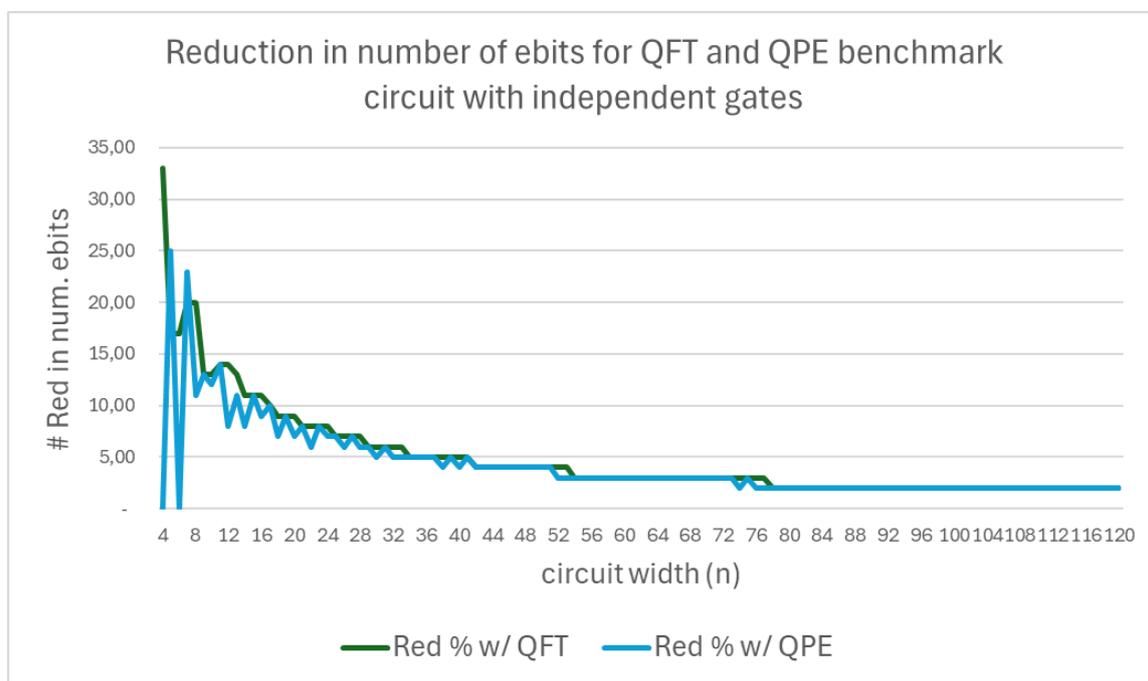

*Figure 15. Impact of hypergraphic partitioning using KL and FM heuristic for AQ and QFT benchmark circuits with independent gates, Source: created by the author (2025).*

In addition to working with the same family of circuits, such as QFT and QPE, in the scenario of independent gates, we have binary and ternary operations; therefore, the impact of the hypergraphic heuristic is complete. In this case, different from what we observed before, the hypergraphic partitioning approach brings benefits for bigger circuits in the number of qubits when compared to the central cutting approach.

For all experiments, we used a machine with 16 GB RAM, a 12th Gen Inter Core i7 with 16 logical processors, and a base speed of 2,20 GHz. The platform for quantum circuits and experiments was based on Python 3 with the Qiskit library for quantum gates and optimization.

## 5. Discussion

Our investigation revealed that employing graph and hypergraphic heuristics such as Kernighan-Lin (KL) and–Fiduccia-Mattheyses (FM) can lead to a minimum cut between partitions equivalent to results from a full circuit. Experiments with spatial partitioning demonstrated a decreasing trend in the reduction of ebits generated by circuit partitioning using hypergraphic heuristics, particularly for Quantum Fourier Transform (QFT) and Quantum Phase Estimation (QPE) circuits. The experiments also presented the same graphics shape for every set of circuits in the spatial cutting approach. When we increased the circuit width, we observed a reduction in the benefits of circuit partitioning, reaching a threshold for min. cut using the hypergraph approach. This is interpreted as a direct relationship between the number of binary and ternary gates for larger circuits, while the same behaviour is observed in the full circuit.

From our experiments with spatial partitioning (gate-cutting), the behaviour of communication reduction for circuit partitioning using hypergraphic heuristics for QFT and QPE circuits shows an overall decreasing trend, starting at 30% for the four qubits and gradually decreasing to 2% for the 120 qubits. However, benchmark circuits, such as Amplitude Estimation (AE) and Quantum Approximate Optimization Algorithm (QAOA), exhibit fluctuations or no benefits from hypergraphic partitioning, especially for circuits with a low Coupling Ratio (Cb).

These observations can be used as thresholds for resource estimators in the future, guiding the best approach to a circuit-partitioning task for a quantum circuit. Considering an initial circuit from a specific family of algorithms, such as QFT, QAOA, VQE, or QPE., we can estimate the impact of circuit partitioning using hypergraphic heuristics against other methods, such as central cutting, and select less expensive strategies for optimization or heuristics, creating space for better reduction of ebits between partitions.

The same conclusion is applicable to the temporal and qubit-wire approaches. Utilizing the dual hypergraphic representation for quantum circuits showed promising results, notably reducing the number of cuts in qubit wires, thus mitigating the impact of segmentation on serial execution. Our study introduces novel insights into circuit dimensions and applications of hypergraphic theory, with the Coupling Ratio shedding light on circuit interconnectivity and aiding in benchmarking cutting processes. Although primal hypergraphs facilitate spatial cutting approaches, dual hypergraphs are more suitable for temporal circuit cutting, leading to significant reductions in communication and initialization costs for scalable quantum computing systems.

This study introduces novel insights into the circuit dimensions and applications of hypergraphic theory for quantum circuit partitioning. Introducing a new dimension termed the Coupling Ratio sheds light on the interconnectivity within circuits, which aids in benchmarking the circuit-cutting process by establishing thresholds. Although the primal hypergraph facilitates spatial cutting approaches (gate-cutting technique), the dual hypergraph derived from it is more suitable for temporal circuit cutting (qubit-wired cutting), leading to significant reductions in communication and initialization costs for scalable quantum computing systems.

Future research will focus on developing a resource estimator tailored for quantum circuit partitioning, building on the efficacy of hypergraphic heuristics, such as KL and FM.

Consolidating new experiments with temporal partitioning and dual hypergraphic representation aims to further elucidate the advantages of this theory, particularly in mitigating the segmentation impact for serial execution. Additionally, investigating the incorporation of weighted hyperedges in both primal and dual hypergraphs for spatial and temporal partitioning will refine our understanding of the factors influencing communication cost reduction between partitions, ultimately optimizing the process, and enhancing system scalability for distributed quantum computing.

Data Access Statement: Research data supporting this publication are available from the IEEE DataPort https://dx.doi.org/10.21227/fqjn-zc55 [38]

## 6. Conclusion

The study effectively demonstrates the potential of hypergraphic heuristics such as Kernighan-Lin (KL) and Fiduccia-Mattheyses (FM) for optimizing quantum circuit partitioning. Key findings include the ability of these heuristics to achieve minimal cuts that mirror those of full circuit implementations, particularly in Quantum Fourier Transform (QFT) and Quantum Phase Estimation (QPE) circuits. However, the benefit of such partitioning decreases with increased circuit width, indicating a scalability threshold. Additionally, while certain circuits like QFT and QPE show decreasing benefits with larger qubit counts, others such as Amplitude Estimation (AE) and Quantum Approximate Optimization Algorithm (QAOA) do not consistently benefit from hypergraphic partitioning, especially when the Coupling Ratio (Cb) is low. The introduction of the Coupling Ratio as a metric for assessing interconnectivity within circuits marks a significant advancement, aiding in the benchmarking and threshold establishment for circuit cutting.

The importance of this study lies in its contribution to the efficient design and optimization of quantum computing systems. By demonstrating the effectiveness and limitations of hypergraphic partitioning, the research provides critical insights that guide the application of quantum circuits in practical, scalable settings. The differentiation between primal and dual hypergraphs for spatial and temporal cutting, respectively, highlights tailored strategies that can significantly reduce communication and initialization costs, pushing forward the scalability of quantum systems.

Looking ahead, future research could focus on several promising areas. Developing a resource estimator tailored for quantum circuit partitioning could enhance the practical application of these findings. Further exploration into the dual hypergraphic representation may provide deeper insights into minimizing segmentation impacts during serial execution. Additionally, incorporating weighted hyperedges in both primal and dual hypergraphs could refine the understanding of how communication costs can be optimized between partitions. Such advancements will continue to optimize quantum computing processes and enhance system scalability, supporting the distributed nature of future quantum computing frameworks.